\documentclass[aps,prb,twocolumn,superscriptaddress,floatfix]{revtex4}

\usepackage{epsfig,graphicx}
\usepackage{natbib}
\usepackage{tabularx}
\usepackage{amsmath}
\usepackage{amsfonts}
\usepackage{amssymb}

\begin{document}

\title{Local order of liquid water at the electrochemical interface}

\author{Luana S. Pedroza}
\affiliation{Department of Physics and Astronomy,  Stony Brook University, Stony Brook, New York 11794-3800, USA}

\author{Adrien Poissier}
\affiliation{Department of Physics and Astronomy,  Stony Brook University, Stony Brook, New York 11794-3800, USA}

\author{M.-V. Fern\'andez-Serra} 
\email[To whom correspondence should be addressed:\\ ]{maria.fernandez-serra@stonybrook.edu}
\affiliation{Department of Physics and Astronomy,  Stony Brook University, Stony Brook, New York 11794-3800, USA}
\affiliation{Institute for Advanced Computational Sciences, Stony Brook University, Stony Brook, New York 11794-3800, USA}

\date{\today}

\begin{abstract}
	We study the structure and dynamics of liquid water
in contact with Pd and Au (111) surfaces using \emph{ab initio} molecular dynamics simulations
with and without van der Waals interactions.
Our results show that the structure of water at the interface of
these two metals is very different.
For Pd, we observe the formation of two different domains of 
preferred orientations, with opposite net interfacial dipoles.
One of these two domains has a large degree of in-plane hexagonal order.
For Au a single domain exists with no in-plane order. For both metals, the structure
of liquid water at the interface is strongly dependent on the use of dispersion forces.
The origin of the structural domains observed in Pd is associated to the interplay
between water/water and water/metal interactions. This effect is strongly
dependent on the charge transfer that occurs at the interface, and which
is not modeled by current state of the art semi-empirical force fields.
\end{abstract}

\maketitle

In the last few years  there has been an explosion of fundamental physics research in the field of electrochemical energy conversion or storage,
driven by the need of optimizing and discovering new materials for renewable energy applications, such as photo-electrochemical fuel cells\cite{MHenderson02,ParrinelloPRL11,GVothJPCC12}.
While both experimental and theoretical surface scientist have made a lot of progress on the understanding
and characterization of both atomistic structures and reactions at the solid/vacuum interface\cite{MichaelidesPRL03,MichaelidesJCP09,AHodgson09,Adrien11,SalmeronScience02,SalmeronPRL04,MichaelidesJCP13,SalmeronPRL04,MichaelidesPRB04,SalmeronJACS09,AHodgson12,GrobNJP09}, the theoretical description of electrochemical interfaces is still lacking behind.
The main reason for this is that a complete and accurate first principles description of both the liquid and the metal interface is still
computationally too expensive. 
Therefore, most studies to date have studied the fully solvated electrochemical interface
using only a classical or semiclassical description. At most, fully first principles studies have only simulated one\cite{SalmeronPRL04,MichaelidesJCP13} 
or two\cite{AHodgson12,GrobNJP09} bilayers
of ordered or semi-ordered water or liquid water structures obtained from classical molecular
dynamics simulations in contact with metal electrodes\cite{YLuoPRB12,GVothJPCC12}. None
of these studies really answers the question of what is the actual structure of pure liquid water at the metal
electrode interface. This is a critical first step to validate current models of the electrochemical, i.e. ionic aqueous solution/metal
interface\cite{Arias2012,DLimmerPNAS13,DLimmerJCP13,DLimmerPRL13}.
%

In this letter we overcome the limitations pointed out by previous studies of the aqueous electrochemical
interface and analyze in detail the structure and interfacial charge redistribution of 
liquid-water interacting with  Pd and Au (111) surfaces at ambient temperature, 
using first principles molecular dynamics. The study will reveal that,
contrary to what was found when studying ice-like water overlayers\cite{MichaelidesPRL11,MichaelidesJCP13}
long range dispersion (van der Waals, vdW) interactions play a critical role in modeling the aqueous/electrode interface
and results are very different to those obtained when vdW interactions are not accounted for. 
Our results will also show that the structure of liquid water at the interface of (111) transition metal
surfaces is not metal independent and that an accurate description of the interfacial chemistry, beyond what is currently described by 
semi-empirical models\cite{Siepmann-Sprik} is needed to simulate the electrochemical interface.


%

%
We perform DFT-based {\it Ab Initio} molecular dynamics (AIMD)  
using the Siesta code\cite{siesta1,siesta2}.
Two different exchange and correlation (XC) functionals are used in this study.
One is the PBE\cite{pbe} gradient-corrected (GGA) functional, which in the past has been
the standard functional choice for water/metal studies\cite{MichaelidesPRL11,MichaelidesJCP13}.
We also perform simulations with a modified version of the functional  proposed by Dion \emph{et al.}\cite{drsll} (DRSLL),
in which PBE is used as the local term of DRSLL functional, 
instead of revPBE as originally proposed.
This vdW-DF$^{PBE}$ functional has recently been shown to produce good dynamical
and structural properties of liquid water\cite{JWangJCP11,mvfs13}.

For Pd(111) surfaces, the unit cell consists of 4 layers of 24 Pd atoms (96 in total) and 80 H$_2$O molecules, 
with size $9.715\times16.826\times26.968$ \AA. This choice reflects
the need of having a unit cell large enough in all directions
to ensure the correct treatment of liquid water. 
No $k$-points were sampled. The in plane
unit cell dimensions ensures an effective sampling of 24-$k$ points. 
%
Norm-conserving pseudopotentials in the Troullier-Martins form\cite{pseudo1} are used to describe core electrons, 
both for metal and water. 
The valence electrons are described using a variationally optimized basis set 
 of numerical atomic orbitals with double-$\zeta$ polarized size. 
More details on the basis sets can be found in Ref. \onlinecite{mvfs04} and \onlinecite{Adrien11}.
%
%

%
Periodic boundary conditions ensure that water is confined at a fixed density
and in contact with the two sides of the metal slab. Therefore two interfaces
are formed in each AIMD.
%
This choice reflects the need of performing comparisons of XC functionals at constant density
instead of having a metal/water/vacuum structure in which the large differences between
equilibrium density of the different functionals would significantly affect the metal/water
interface\cite{JWangJCP11}.
To evaluate the importance of sampling and of the initial configuration of the water/metal structure,
 we have performed two distinct simulations with different initial conditions for the water molecules.
This sampling part of our study is only done for the PBE simulations, although short simulations
using vdW-DF$^{PBE}$ provide the same qualitatively conclusions. 
%
In one simulation, two 2-D monolayers of ice of 16 water molecules each were adsorbed at each side of the Pd-(111) slab. 
 48 additional liquid water molecules were inserted in the remaining space between the two Pd surfaces.
The second one contained the same total number of H$_2$O molecules (80),  all of them in liquid form
inserted between the two metallic slabs. 
The motivation behind this choice was, in addition to evaluate the importance of initial conditions,
to test whether the well characterized 2-D ice bilayer\cite{MichaelidesPRL11,MichaelidesJCP13}
would retain certain amount of order, or even induce some additional order when the metal surface 
is fully wet with liquid water at room temperature.
While a previous work\cite{GrobNJP09} has studied the stability of up to two 2-D ice bilayers
in different metals, no study has yet considered their stability in fully solvated
conditions at room $T$. Our work offers an answer to this question.
All our production runs are at least 10 ps long, with a time step of 0.5 fs.
They are performed in the micro-canonical ensemble, and
follow an AIMD annealing run to $T=325$ K for PBE and $T=300$ K for vdW-DF$^{PBE}$  during 5 ps.
Our results from the AIMD show that the different initial conditions do not appear to affect the final interfacial
water structures, the results being qualitatively similar for both cases, indicating that 10 ps are enough
to achieve the formation of a converged interfacial water structure. In deed, in all our simulations,
the interfacial structure formed within 2 ps, after the initial 5 ps pre-equilibration period.
Results comparing these two simulations and a table with all the simulations performed in
this study are presented in the Supplementary Information\cite{SI}. 
%
%

In Fig.\ref{densityZ}(a) and (c) we show the probability
of finding O and H atoms at a given $z$ distance from the surface at the two interfaces  
for both PBE  and  vdW-DF$^{PBE}$, respectively. 
The position of the last Pd(111)  layer of atoms at the Pd(111) surface is set to zero in $z$ for both sides of the slab.

\begin{figure}[!htb]
\begin{center}
\includegraphics*[width=\columnwidth]{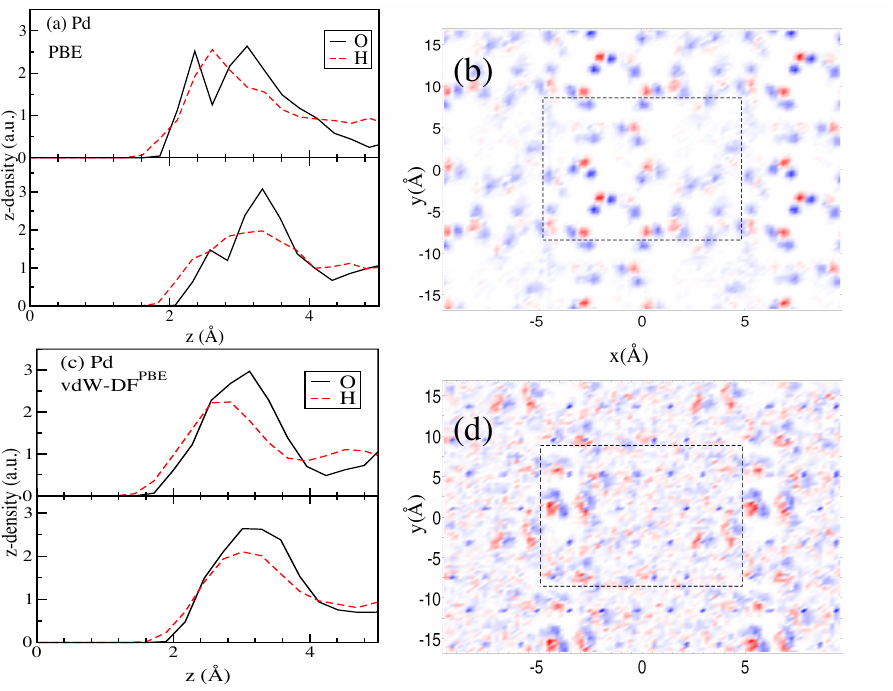}
\caption{Atomic densities for O and H atoms computed along the $z-$distance (in \AA) at the two Pd/H$_2$O interfaces for
 (a) PBE and (c ) vdW-DF$^{PBE}$ .
The position of the last layer of atoms at the Pd(111) surface is set to zero in $z$ for both sides of the slab. 
The $XY$ surface probability at the interfaces is shown in (b) 
for PBE and in (d) for vdW-DF$^{PBE}$ . 
Water molecules can be seen formed by O (red) and H (blue). The first Pd layer below is 
shown in black spots. 
The square indicates the unit cell used in the AIMD simulations.}
\label{densityZ}
\end{center}
\end{figure}

For PBE, the interface in the upper plot presents a sharp double peak for the O atoms.
The peak closer to the surface originates from molecules
adsorbed in a ``flat'' geometry, i.e. the molecule's dipole moment is mostly parallel to the surface. 
The second peak is due to ``up/down" molecules, which donate one H-bond
to a flat molecule and the other to a molecule in a layer above (``up'')
or to the metal (``down''). 
This formation resembles the 2D ice structure\cite{SalmeronPRL04,SalmeronScience02,Adrien11}, 
where a monolayer of water covers a metallic surface with 
a hexagonal arrangement of flat and up(down) water molecules. 
Fig.\ref{densityZ}(b) shows the probability of finding an atom at a particular position on the $XY$ plane, 
integrated over the first layer and averaged over all the simulation length. 
The hexagonal lattice of surface Pd atoms is shown 
in black small spots.
Oxygen atoms are shown in red and H atoms in blue.
At this interface we observe the
formation of an ordered domain of hexagons and 
pentagons, with very stable
flat molecules.

The other interface, shown in the bottom of Fig.\ref{densityZ}(a)  presents
a smaller peak related to the presence of flat molecules close to the surface and 
a broad peak composed of mostly ``down" molecules.
In this case there is little in-plane order, with a structure very similar to what is shown in Fig.\ref{densityZ} (d).
We refer to this as a ``disordered" domain.
The complete $z$- and $XY$-density profiles for all the simulations are provided in the SI.

When the simulation is repeated using vdW-DF$^{PBE}$ for XC, the interfacial order changes substantially.
The double peak in the density profile is not well defined at any interface, 
as shown in Fig.\ref{densityZ}(c). 
Flat orientations, which are driven by donor-covalent, H-bond type interactions between
one of the O lone pairs and the metal\cite{MichaelidesJCP09,Adrien11}, are less favored 
when vdW interactions are accounted for. These tend to favor more perpendicular orientations,
which are also favored by  dipole-image-dipole electrostatic interactions.\cite{Adrien11}
In addition, as already discussed in other studies\cite{JWangJCP11,mvfs13},
dispersion interactions largely increase the room temperature  diffusivity of
vdW-DF$^{PBE}$ liquid water with respect to PBE.
Indeed, the interfacial profiles in Fig.\ref{densityZ} are integrated for the full simulation length.
A time dependent analysis of these profiles reveals that the water layers  at the interfaces in the vdW
simulation fluctuate between the previously described ordered and disordered
domains.
At any particular time there is always an asymmetric interfacial configuration, resulting
in a more disordered structure on average, as shown in Fig.\ref{densityZ}(d).

\begin{figure}[!htb]
\begin{center}
\includegraphics*[width=\columnwidth]{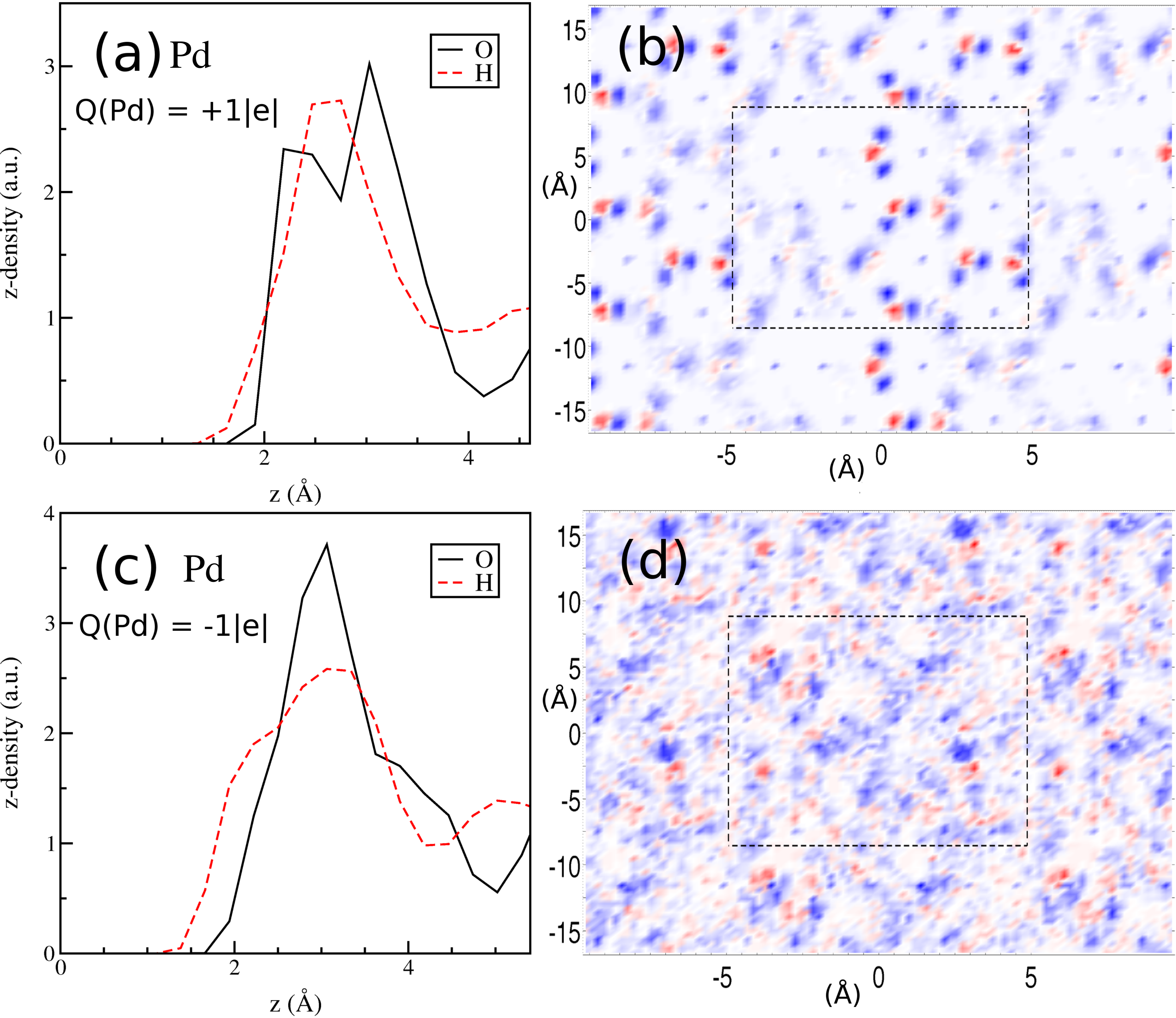}
\caption{Atomic densities for O and H atoms computed along the $z-$distance (in \AA) at the  Pd/H$_2$O interface for
charged simulations with net charge +1$|e|$ in (a) and -1$|e|$ in (c ) using PBE XC. 
Slab surface correspond to the zero in $z$. 
The $XY$ surface probability at the interfaces is shown in (b) 
for  +1$|e|$ and in (d) for -1$|e|$.
Water molecules can be seen formed by O (red) and H (blue). The first Pd layer below is 
shown in black spots. 
The square indicates the unit cell used in the AIMD simulations.}
\label{densityZcharged}
\end{center}
\end{figure}

%
%
Depending upon the orientation of the water molecule towards the surface
the metal may screen a negative or positive charge. 
%
%
When the O atom is oriented towards the surface, it induces a positive image charge
near the metal surface. 
When the molecular orientation has the H atom facing the metal, the opposite occurs and
a negative image charge is induced.
Besides this screening effect, the donor-covalent type of bond between 
flat molecules and the metal results in a charge transfer from the
 water molecules to metal\cite{Adrien11}.
We can then associate to regions with more flat molecules a domain
where the surface will be more positively charged. 
And, contrary, a negative domain for regions where there is a presence
of more molecules with hydrogen pointing down.
The idea of charged domains has been shown before for water monolayers.\cite{DoubletCT13}
We can compute the net water to metal charge transfer per configuration, by integrating the charge density
difference between the solvated metal and the metal and water regions isolated. Assuming
that most of this charge is transferred by `flat" molecules\cite{Adrien11}, 
we estimate that for PBE the ordered domain 
shows a net 0.004  $e^-$/\AA$^2$ charge loss to the metal. The disordered domain
transfers only 0.002 $e^-$/\AA$^2$. In this estimation we are assuming a linear relation
between the charge transfer and the number of flat molecules. 

We have  performed two additional AIMD simulations with a positive (+1$|e|$) and 
negative (-1$|e|$) extra net charge to analyze how the domain order is
modified under charged electrode conditions.  Each simulation is also 10 ps long. PBE was chosen for XC and
the unit cell is the same as for the neutral case.
%
%
The density profiles for these charged systems are shown in Fig. \ref{densityZcharged} (a)-(d). 
For the system with a positive net charge, we observe the formation of the double peak in both interfaces, 
similar to the previously described ordered domain. 
On the other hand, when the system has a negative net charge, the number of
molecules at both interfaces with down orientations increases, and the disordered domain is formed at the two interfaces.

%

\begin{figure}[!htb]
\begin{center}
\includegraphics*[width=\columnwidth]{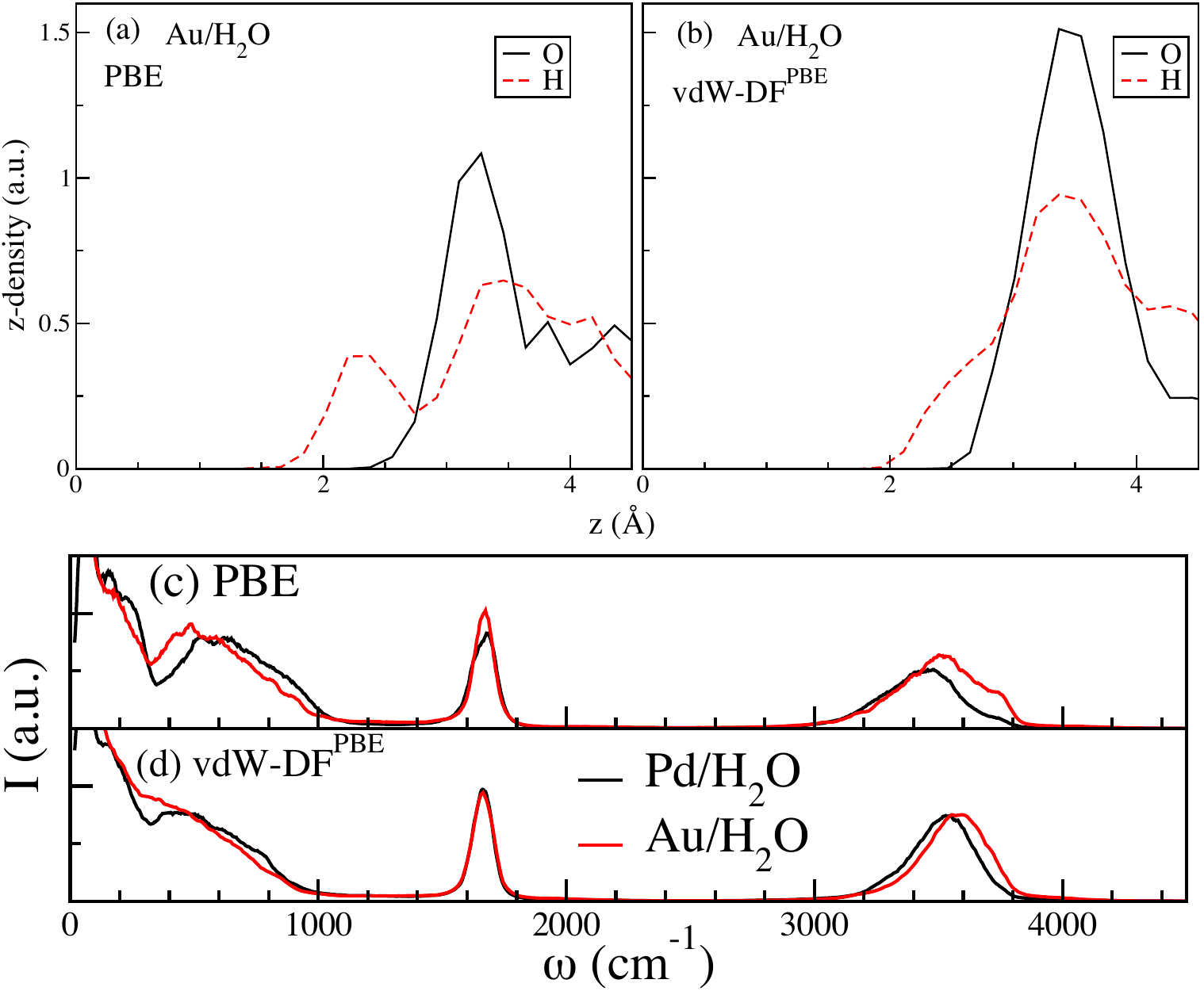}
\caption{Top: atomic densities for O and H atoms computed along the $z-$distance (in \AA) at the  Au/H$_2$O interface for
(a) PBE and (b) vdW-DF$^{PBE}$. The surface of the slab is set to  zero in the $z$-axis. 
Bottom: vibrational density of states for Pd/H$_2$O (black lines) and  Au/H$_2$O (red lines)
as obtained for (c) PBE and (d) vdW-DF$^{PBE}$ simulations.
}
\label{ZDensAu}
\end{center}
\end{figure}

%
Current water/metal interatomic potentials\cite{Siepmann-Sprik,DLimmerPNAS13,DLimmerJCP13} do not take into account 
the charge transfer effect which seems to be the origin of the domains. They also were parametrized to reproduce 
single molecule water absorption geometries, which we have shown do not reproduce the actual interfacial
geometry of fully solvated surfaces. However they describe very accurately the
electrostatic interaction between the image charges at the metal and the solvent.
To compare our results with a system where water/metal interactions are more electrostatically
dominated, due to the absence of charge transfer\cite{GrobNJP09}, we  have also performed  AIMD simulations
(with both PBE and  vdW-DF$^{PBE}$ functionals) of the Au(111)-H$_2$O interface\cite{Details_simulations}.
The $z$-density profiles are shown in Fig. \ref{ZDensAu}(a,b).
Contrary to the Pd-H$_2$O system, we do not observe a double peak at the interface at any point, indicating that 
with Au, there are more down/up molecules at the interfaces. Interestingly, in this
case, both functionals produce similar interfacial order, although they present very different 
vibrational density of states (DOS), also shown in Fig. \ref{ZDensAu}(c). 
These DOS spectra provide information about the hydrophobic nature of the interfaces,
given that hydrophobic interfaces are characterized by the presence of dangling, non H-bonded 
molecules, with a blue-shifted peak
in stretching region of the spectrum\cite{Galli08}.  
This dangling OH signature
is only weakly present in the Au/PBE simulation, indicating that in general
these interfaces are less hydrophobic than originally claimed using
classical potentials\cite{DLimmerPNAS13}.
However, the relative blueshift of the overall stretching peak from Pd to Au
indicates that the H-bond network is weakened in water at the interface of Au,
which can also be taken as an indication of larger hydrophobicity for Au with 
respect to Pd.

Our results with Pd and Au allow us to explore changes in the work function of these metals
upon solvation, and to make connections to experimental results.
A standard concept is electrochemistry is the potential of zero charge (PZC).
This is the electrode's potential at the so called point of zero charge, which is the state of the pure solvent, free of ions.
 In principle this is the same as the work function ($\Phi$)  of the solvated electrode.\cite{Tripkovic11}
Here we evaluate change of the
work function of the metal upon solvation, $\Delta\Phi=\Phi_{m/H_2O}-\Phi_{m}$ (m=Au/Pd).
When the metal surface is fully solvated this change does not
necessarily need to be the same as the change observed upon gas absorption\cite{Tripkovic11,YLuoPRB12}.
 In the latter case the change is due to the adsorption of one or very few water monolayers with 
very specific structures, and in general
$\Delta\Phi<0$, indicating that the dipole of the adsorbed water molecules is antiparallel to the
clean surface's dipole (which is always perpendicular to the surface and pointing inwards).
 Therefore in those cases water adsorbs with the oxygen atom facing the surface, as already shown
in numerous studies\cite{MichaelidesPRL03,MichaelidesJCP09,SalmeronScience02,SalmeronPRL04,Tripkovic11}.
For fully solvated surfaces, obtaining an accurate $\Delta\Phi$ is a challenge
both theoretically\cite{YLuoPRB12} and experimentally\cite{PollackCPL12,Tripkovic11}. 
Although some works have analyzed the liquid-water/metal interface,
they all used semi-empirical force fields, some of which
include the polarizability of the metal\cite{Siepmann-Sprik}, but do not account for water polarizability 
and charge transfer processes. As a consequence, these models are not accurate enough to
 reproduce the complex liquid/metal interactions at the interface and they cannot
be used to predict the exact structure of the water/metal interface, not at least
without validation from first principle simulations.
 

%
To compute $\Phi$ for the bare surface ($\Phi_{m}$, m=Au/Pd)
we evaluate $\Phi=E_v-E_F$, where
$E_v$ is the planar averaged electrostatic potential energy in the near vacuum region next to the bare surface
and $E_F$ is the Fermi level of the system.
The solvated surface's ($\Phi_{m/H_2O}$, m=Au/Pd) work function is computed in the same way, after
correct alignment of the metal's electrostatic potential with that of the bare surface, therefore referencing
with respect to the same vacuum level.
We computed $\Phi$ for representative configurations  from the AIMD.
In table \ref{table1} we show the time-averaged work function change, $\Delta\Phi$ and $\Phi_m$,  for both Pd and Au.

\begin{table}
\caption{\label{table1} First column, work function $\Phi$(eV) of bare (111) Pd and Au as obtained using different XC potentials.
In parenthesis, the experimental values for the bare (111) surfaces work functions are provided.
Second column, averaged  work function change ($\Phi_{metal/H_2O}-\Phi_{metal}$) (eV) as obtained from our AIMD simulations.
 Last column, estimation of net charge transfer from water to the metal interface, in $e^-/$\AA$^2$}
\begin{tabularx}{\columnwidth}{c|lXXX}
 \hline\hline
 & XC & $\Phi$(eV) & $\Delta\Phi$(eV)& $\sigma_{CT}$($e^-/$\AA$^2$) \\
  \hline
 Pd & PBE           & 5.15 (5.6\cite{MichaelsonJAP77}) & 1.15 $\pm0.44$ & 3.0 $\cdot10^{-3}$\\
 Pd & vdW-DF$^{PBE}$& 5.40  & 1.22 $\pm0.28$ & 1.5 $\cdot10^{-3}$\\
 \hline
 Au & PBE           & 5.13 (5.31\cite{MichaelsonJAP77})  & -0.58 $\pm0.37$ & 8.0 $\cdot10^{-5}$\\
 Au & vdW-DF$^{PBE}$& 5.34  & -0.68 $\pm0.39$ & 3.0 $\cdot10^{-3}$\\

 \hline\hline
\end{tabularx}
\end{table}

%
%
The work function of the Pd/H$_2$O interface increases with respect to the Pd(111)
clean surface. The two domains previously described, in-plane ordered and in-plane disordered
have opposed interfacial dipoles. The disordered domain, with a larger number of ``down'' molecules,
has a net negative interfacial dipole moment, i.e. with the dipole vector antiparallel to
the vector normal to the surface. This is the reason why the net work function change is positive,
indicating that, despite the ordered domain would tend to decrease the work function both
because of the charge transfer and ``push-back" effect\cite{BraunAdvMat2009} associated to the physisorption
of flat molecules,
the large interfacial dipole of the disordered domain dominates the sign of the work function change.
It is interesting to note that with vdW-DF$^{PBE}$, the net charge transfer is reduced, which correlates
with an increase of the solvated work function, as expected.

For Au the change in work function is smaller and of opposite sign as compared to Pd. 
 The net interfacial dipole is small, because most of the interfacial water are flat-down or flat-up, therefore
canceling each others contribution to the net dipole. Overall the ``push-back'' effect\cite{BraunAdvMat2009}  dominates
and the work function decreases.
 We observe a negligible amount of charge transfer for Au, as we expected.
Overall our results show that the change in work function upon solvation is strongly metal dependent.

%
%
%
%
%
%
%
%


%




In conclusion, we have showed how liquid-water interacts with Pd and Au (111) surfaces
using \emph{ab initio} molecular dynamics simulations. 
Our results show that the inclusion of the electronic degrees of freedom (in this case, via DFT) is essential to 
properly describe the interaction of water and metal surfaces in general.
This study has explored two important aspects, critical for understanding how to achieve accurate simulations
of the electrochemical interface.
First of all, we have evaluated the importance of vdW interactions in DFT when simulating
liquid water and metallic surfaces.
Our results show that, in order to reproduce the interfacial liquid structure, accounting for vdW
interactions is particularly critical, mostly because of the description
of liquid water itself.
Secondly, we have addressed whether the nature of the metallic surface affects the liquid structure
at the interface. 
Our results show that, contrary to what has been claimed using classical potentials\cite{DLimmerPNAS13},
the order of water at the interface strongly depends on the metal under consideration and is not always hydrophobic.
In particular, for Pd, we have shown that liquid water adsorbs forming two different domain structures,
which are driven by the strong physisorption of ``flat''-type water molecules and
associated charge transfer mechanism. 
For Au, we do not observe the formation of domains, which is related to the lack of charge transfer
at the interface.

We acknowledge fruitful discussions with Alexandre R. Rocha.
This research used computational resources at the Center for Functional Nanomaterials, Brookhaven
National Laboratory, which is supported by the US Department of Energy under
Contract No. DE-AC02-98CH10886. The work was funded by DOE Early Career Award No. DE-SC0003871.

\bibliography{pdH2O}

\begin{thebibliography}{37}
\bibitem[{Det()}]{Details_simulations}
\bibinfo{note}{The Au simulations were performed for a unit cell in the XY
  direction half the size of that of Pd, with area=$9.753\times9.373$\AA\, and
  40 water molecules in total.}
\end{thebibliography}

\end{document}